\documentclass[fleqn,usenatbib]{mnras}

\usepackage{newtxtext,newtxmath}

\usepackage[T1]{fontenc}

\DeclareRobustCommand{\VAN}[3]{#2}
\let\VANthebibliography\thebibliography
\def\thebibliography{\DeclareRobustCommand{\VAN}[3]{##3}\VANthebibliography}

\usepackage{graphicx}	
\usepackage{amsmath}	
\usepackage{array}

\usepackage{booktabs}
\usepackage[dvipsnames]{xcolor}

\newcommand{\skm}{\mathrm{km^{-1}\,s}}
\newcommand{\Mmax}{M_{\rm UV}^{\rm max}}
\newcommand{\halfspace}{\hspace{1pt}}
\newcommand\HI{{\hbox{H\halfspace$\rm \scriptstyle I$}}}

\newcommand\HeII{{\hbox{He\halfspace$\rm \scriptstyle II$}}}
\newcommand\HeIII{{\hbox{He\halfspace$\rm \scriptstyle III$}}}

\newcommand\Lya{Lyman-$\alpha$}

\def\orcid#1{\href{https://orcid.org/#1}{\includegraphics[keepaspectratio,width=0.7em]{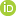}}}

\newcommand{\orcidauthorA}{\orcid{0000-0002-5445-461X}} 

\newcommand{\orcidauthorB}
{\orcid{0000-0002-5451-9057}} 

\title[Flux $P(k)$ constraints on QSO-assisted reionization]{\Lya\ forest 1D flux power spectrum constraints on QSO-assisted reionization models}

\author[Meiksin \& Ir\v{s}i\v{c}]{
Vid Ir\v{s}i\v{c}$^{1,2}$\thanks{E-mail: v.irsic@herts.ac.uk (VI)}\,\orcidauthorA,
Avery Meiksin$^{3}$\thanks{E-mail: meiksin@ed.ac.uk (AM)}\,
\orcidauthorB
\\
$^{1}$Center for Astrophysics Research, University of Hertfordshire, College Lane, Hatfield AL10 9AB, UK\\
$^{2}$Department of Physics, Astronomy and Mathematics, University of Hertfordshire, College Lane, Hatfield AL10 9AB, UK\\
$^{3}$Institute for Astronomy, University of Edinburgh, Blackford Hill, Edinburgh EH9 3HJ, UK\\
}

\date{Accepted XXX. Received YYY; in original form ZZZ}

\pubyear{\the\year{}}

\begin{document}
\label{firstpage}
\pagerange{\pageref{firstpage}--\pageref{lastpage}}
\maketitle

\begin{abstract}
Recent JWST observations have revealed a large population of faint Quasi-Stellar Objects (QSOs) at redshifts $4<z<6$, including a highly reddened sub-population of Little Red Dots. We exploit the sensitivity of the \Lya\ forest 1D flux power spectrum to thermal fluctuations in the Intergalactic Medium (IGM) to place statistical constraints on the spectral properties of the QSOs. By post-processing simulated \Lya\ forest spectra from the Sherwood-Relics suite of \Lya\ forest simulations with added \HeII\ reionization by QSOs, in conjunction with published precision measurements of the \Lya\ forest 1D flux power spectrum between $4.2<z<5.0$, we find that the addition of temperature boosts of the IGM within the \HeIII\ regions improves agreement with the measured power spectra. While the contribution of the Little Red Dot population to \HeII\ reionization is consistent
with both a mild temperature boost of $\Delta T_b=1\times10^4$~K for soft spectra QSOs and $\Delta T_b=2\times10^4$~K for hard spectra QSOs, a contribution from the larger population of faint QSOs found by JWST having $M_\mathrm{UV}>-21.6$ is excluded at the $2\sigma$ level if their spectra are sufficiently hard to boost the IGM temperature in \HeIII\ regions by $\Delta T_b=2\times10^4$~K or greater.
\end{abstract}




\section{Introduction}
\label{sec:intro}

The structure of the Intergalactic Medium (IGM) is governed primarily by the growth of cosmic density perturbations. Cosmological numerical simulations combining gravity and hydrodynamics reproduce a wide range of statistical properties of the IGM as quantified through measurements of the \Lya\ forest, absorption features detected in the spectra of bright background Quasi-Stellar Objects (QSOs). As such, \Lya\ forest observations have been used to constrain cosmological models and properties of dark matter. \citep[See][for reviews]{Meiksin09, mcquinn16}.

Secondary influences on the IGM arise from the impact of photoionization sources on the thermal and ionization state of the IGM. These effects are detectable using high resolution, high signal-to-noise ratio measurements of the  \Lya\ forest. They become the limiting factors in constraining dark matter properties using high resolution data, able to probe length scales comparable to the Jeans length of the IGM. The \Lya\ forest statistics reveal the impact of the first photoionizing sources as the hydrogen in the IGM becomes reionized at high redshifts, $z>5$, most likely by galaxies and QSOs \citep{2006AJ....132..117F, 2015MNRAS.447.3402B, Boera2019, Kulkarni2019b, Onorbe2019, Puchwein2023}, and helium becomes fully ionized at moderate redshifts of $2<z<5$ by QSOs \citep{MM94, 2000ApJ...534...57B, MBM01, Tittley07, 2011ApJ...726..111S, 2015MNRAS.450.4081P, 2017MNRAS.465.2886D, 2019ApJ...875..111W}.

The primary effect of the QSO sources on the hydrogen is through the production of large-scale inhomogeneities in the temperature field of the IGM on scales of tens of comoving megaparsecs through the generation of photoionized \HeIII\ regions, with the temperature boosted by as much as $1-2\times10^4$~K \citep{1997ApJ...475..429M, 1999ApJ...520L..13A, 2004MNRAS.348L..43B, 2009ApJ...694..842M, 2013MNRAS.435.3169C, 2017MNRAS.468.3718K, 2023MNRAS.519.5743L, 2024MNRAS.532..841B}, resulting in large-scale modulations of the hydrogen radiative recombination rate and the \HI\ fraction. A secondary effect is to create spatial inhomogeneity in the \HI\ photoionization rate, although spatial correlations in the radiation field are expected to be small on scales exceeding a few to several comoving megaparsecs at $z<5$ \citep{2020MNRAS.491.4884M}.

At high wavenumbers, $0.01<k<0.1$~s\ km$^{-1}$, the hydrogen reionization history introduces an uncertainty in the 1D flux power spectrum of about 10 percent at $z>4$, comparable to the measurement error \citep{Puchwein2023}. The uncertain reionization history accordingly becomes a limiting factor in constraining dark matter particle masses and the neutrino mass.

At redshifts $z<5$, predictions for the flux power spectrum are also limited by uncertainty in QSO numbers and spectra, as QSOs are the only known sources able to photoionize the singly to doubly ionized helium in the IGM. The elevated temperature in the \HeIII\ regions they produce suppresses the flux power at high wavenumbers by amounts that depend on their numbers, spectral hardness and lifetimes \citep{2024MNRAS.535.1035M}.

The number density and EUV spectra of QSOs are uncertain at high redshifts. From a comprehensive analysis of colour-selected QSOs from several UV-optical surveys, \citet{2019MNRAS.488.1035K} derived three separate models for the QSO luminosity function. For QSOs at $z>4$, these describe the bright-end, with QSO absolute AB magnitues at restframe 1450~A in the range $-30<M_\mathrm{UV}<23$. The x-ray selected sample of \citet{2019ApJ...884...19G} shows a large population of QSOs in the magnitude range $-22<M_\mathrm{UV}<-19$ at $z\sim5.6$. Allowing for this population helps to recover the broad \HI\ optical depth distribution found by \citet{Bosman2022} \citep{2024MNRAS.535.1035M}. An even larger population of faint ($M_\mathrm{UV}^\mathrm{AB}>-21$) QSOs at $4<z<6$ has been identified in JWST images \citep{2023ApJ...959...39H, 2024A&A...691A.145M}. Recent work suggests a population of faint QSOs is consistent with the Lyman-$\alpha$ opacity cumulative distribution function measurements for \HI\ \citep{Asthana2025b} and \HeII\ \citep{2024MNRAS.532..841B}. \citet{Asthana2025b} moreover highlight that a small faint QSO contribution to \HI\ reionization better explains the high optical depth tail of \citet{Bosman2022}. The degree to which \HeII\ reionization by the JWST population will suppress the 1D flux power spectrum at high wavenumbers has not been quantified. We do so in this paper, and infer what limits the 1D flux power spectrum may place on the properties of the population.

In the next section, we summarise the IGM simulations we use. In Sec.~\ref{sec:QSOmodels}, we describe the QSO modelling used to post-process the simulations. We then describe the data and statistical analysis. In Sec.~\ref{sec:results}, we present our results, before ending with the Conclusions.

\section{Methodology}
\label{sec:method}

\subsection{Simulations}
\label{sec:sims}

This study uses a suite of simulations from the Sherwood-Relics project \citep{bolton17,Puchwein2023}. These are high-resolution, cosmological volume, hydrodynamical simulations suited for the study of the IGM and the Lyman-$\alpha$ forest. Full details may be found in \citet{Molaro2022} and \citet{Puchwein2023}; here we summarize the main points for brevity.

The simulations used a customized version of {\tt P-Gadget3} code \citep{springel05} for simulated box sizes of 20 $\mathrm{cMpc}/h$ and $2\times 1024^3$ dark matter and baryon particles, respectively. The simulations extend to $z=4$, and the box size and particle number were chosen to resolve the small scale structure of the IGM \citep{lukic15,Doughty2023}. For the purpose of correction for the numerical resolution additional models were used with $10\;\mathrm{cMpc}/h$ boxsize and $2\times 512^3$ and $2\times 1024^3$ particle numbers, following \citet{Irsic2024}. All the models use a simplified and efficient star formation prescription called {\tt Quick\_lya} \citep{Viel04}, where gas particles are converted into collisionless stellar particles upon reaching overdensity and temperature thresholds ($\Delta > 10^3$, $T < 10^5\;\mathrm{K}$). All simulations assume a flat $\Lambda$CDM cosmology with $\Omega_\Lambda = 0.692$, $\Omega_m = 0.308$, $\Omega_b = 0.0482$, $\sigma_8 = 0.829$, $n_s = 0.961$, $h=0.678$ and a primordial helium mass abundance of $Y_p = 0.24$.

A subset of the Sherwood-Relics suite of simulations spans 12 different thermal histories, with varied photoheating and photoionization rates of a uniform UV background synthesis model \citep{Puchwein19}. Thermal histories were labeled with the cumulative heat injected in each of the observed redshift bins ($u_0$) \citep{Nasir16,Boera2019,Irsic2024}. Following previous work \citep{Boera2019,Gaikwad20,Irsic2024}, each of these simulations was post-processed by rotations and translations in the temperature-density plane to obtain a uniform grid of $10 \times 10$ in the gas parameters of the temperature at mean density ($T_0$) and the slope of the temperature-density relation ($\gamma$). Finally, in the post-processing step, for each of the $12 \times 10 \times 10$ simulations the value of the mean transmission was additionally varied across a uniform grid of 15 points around the default redshift evolution of the effective optical depth $\tau_{\rm eff}$ adopted from \cite{Boera2019}. This resulted in $12 \times 10 \times 10 \times 15$ models for each cosmology, varying the amplitude of the $\Lambda$CDM matter clustering $\sigma_8 = [0.754, 0.804, 0.829, 0.854, 0.904]$ and adiabatic spectral index $n_s = [0.921, 0.941, 0.961, 0.981, 1.001]$.

\subsection{QSO-assisted reionization model}
\label{sec:QSOmodels}

Following the approach of \cite{2024MNRAS.535.1035M}, we assess the impact of \HeIII\ regions using the output of the $10\,h^{-1}$~cMpc $2\times1024^3$ simulation of \cite{Puchwein2023}. \HeIII\ regions are produced by QSOs drawn randomly from Models 1 and 3 of the \cite{Kulkarni2019} QSO luminosity function with a minimum AB absolute magnitude at rest frame 1450A $M^\mathrm{min}_\mathrm{UV}=-30$ and varying upper magnitudes $M^\mathrm{max}_\mathrm{UV}$. An extrapolation of Model 1 to faint magnitudes ($M_\mathrm{UV}<-18$) closely matches the number of faint AGN detected by JWST \citep{2023ApJ...959...39H, 2024A&A...691A.145M}, with, for the magnitude range $-21<M_\mathrm{UV}<-18$, an average AGN number density of $1.4\times10^{-4}\,\mathrm{cMpc}^{-3}$ between $4<z<6$. By contrast, the number density averaged over $4.2<z<5.5$ for the same magnitude range from Model 3 is $1.5\times10^{-5}\,\mathrm{cMpc}^{-3}$, matching the number density of Little Red Dots (LRDs) \citep{2024Natur.628...57F, 2024ApJ...964...39G, 2024ApJ...963..129M}. A QSO lifetime of 30~Myr is assumed in the computations; results are not very sensitive to this choice.

QSO positions are chosen randomly in extended boxes outside the simulation volume, allowing for QSO \HeIII\ regions to grow both within and into it from the outside. The IGM temperature within the \HeIII\ regions is boosted by $10^4$~K, a typical expected value, or $2\times 10^4$~K for the case of a hard spectrum ($L_\nu\sim\nu^{-0.5}$) \citep{2023MNRAS.519.5743L}. Past boxes along a light cone are included to follow the evolution of the growing \HeIII\ regions. This allows for the inclusion of relic, recombining \HeIII\ regions after the QSO dies. Multiple random realisations of the QSO population are generated to converge on their effect on the \HI\ \Lya\ forest power spectrum.

\begin{figure*}
	\includegraphics[width=\textwidth]{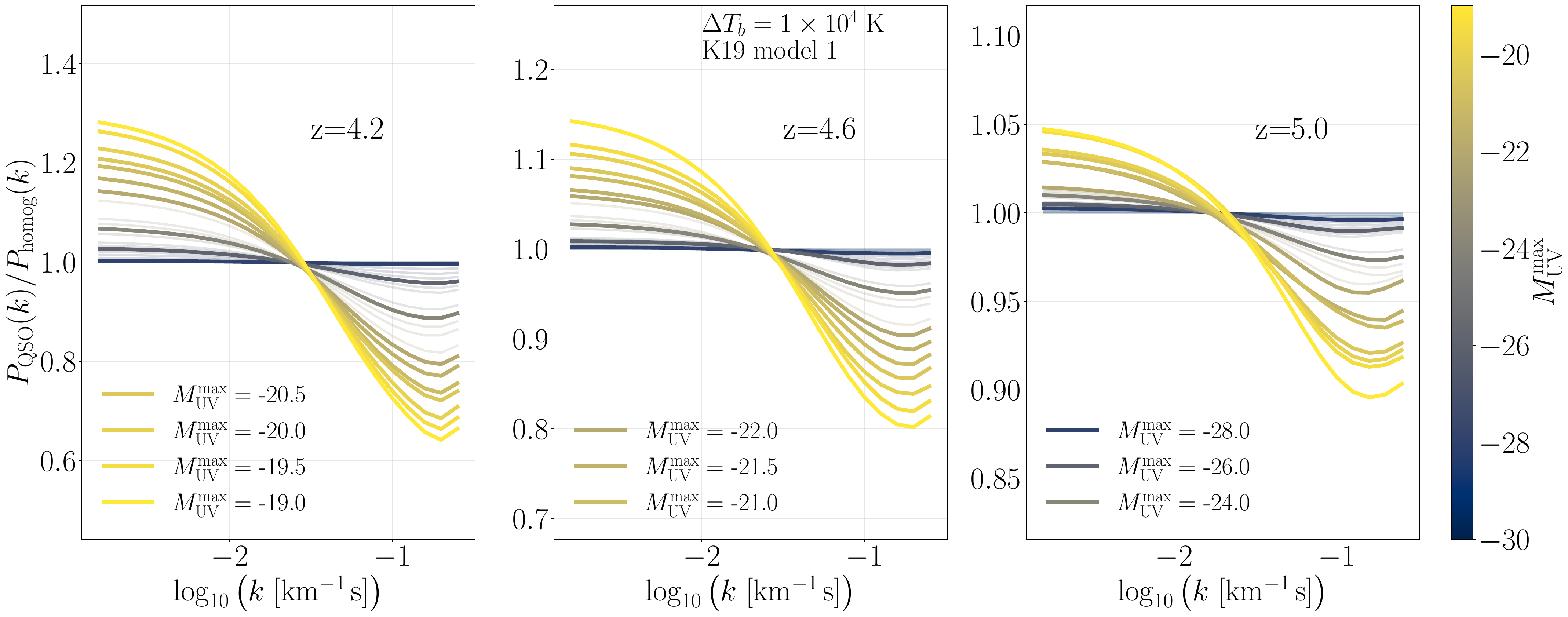}\\
    \includegraphics[width=\textwidth]{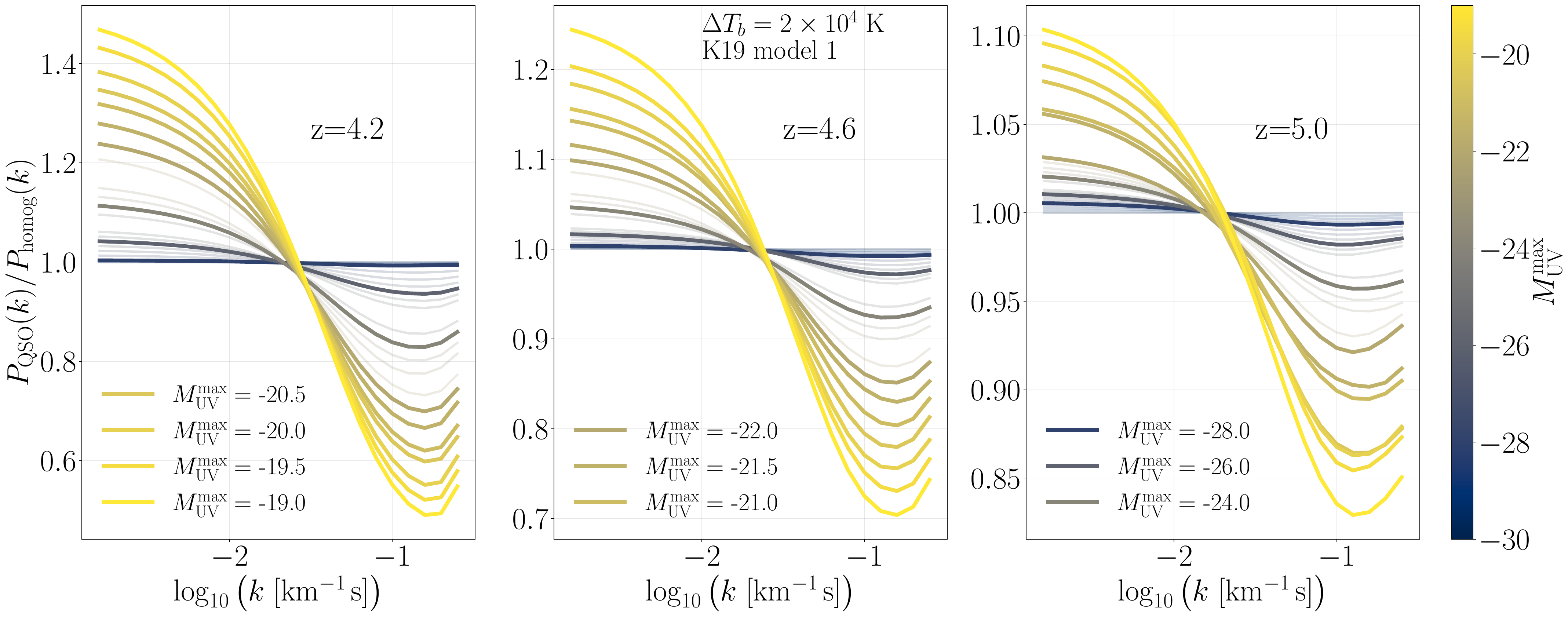}\\
    \includegraphics[width=\textwidth]{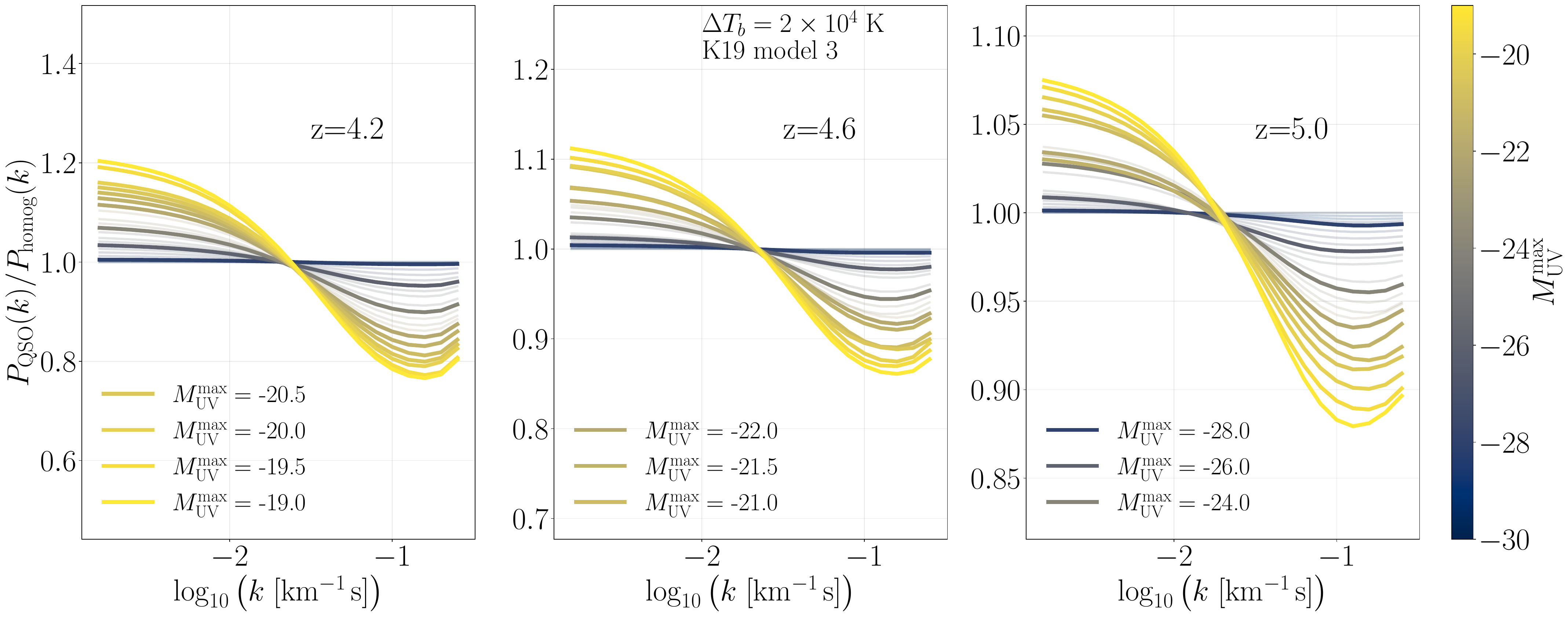}\\
    \caption{The ratio of flux power spectra with and without QSO assisted modelling for a temperature boost within \HeIII\ regions of $\Delta T_b=2\times10^4$~K ({\it middle}, {\it bottom} row) or $\Delta T_b=1\times10^4$~K ({\it top} row). Different coloured lines correspond to different $M_{\rm UV}^{\rm max}$ up to which the QSO luminosity function was integrated. Larger suppression results when integrating down to fainter objects. The results show the output of the likelihood emulator, with the original models (in bold) extrapolated below $M_{\rm UV}^{\rm max}=-22$. The assumed limit is $M_{\rm UV}^{\rm max}=M_{\rm UV}^{\rm min}=-30$ which returns the homogeneous reionization model. The models were computed using the \citet{Kulkarni2019} QSO luminosity function Model 1 ({\it top}, {\it middle} row) or Model 3 ({\it bottom} row), with quasar lifetime $\tau_Q=30\;\mathrm{Myr}$ and integrated from the minimum magnitude of $M_{\rm UV}^{\rm min}=-30$.
      \label{fig:p1d_qso_emulator}}
\end{figure*}

The resulting effects on the 1D flux power spectra are shown for $\Delta T_b=2\times10^4$~K in Fig,~\ref{fig:p1d_qso_emulator} (bold curves), The power at $k>0.01$~s\ km$^{-1}$ is suppressed by the elevated temperature in the \HeIII\ regions, while the suppressed power is compensated by boosts at smaller wavenumbers. Also shown in Fig,~\ref{fig:p1d_qso_emulator} are the results from an emulator that extrapolates the suppression for $M_\mathrm{UV}^\mathrm{min}\rightarrow -30$, corresponding to the limiting case of no QSO-produced \HeIII\ regions.

\subsection{Data}
\label{sec:data}

Data anaylsis of this study uses the published 1D flux power spectrum measurements of \cite{Boera2019}. These measurements were obtained from Keck/HIRES and VLT/UVES high-resolution observations of 15 quasars in each of three redshift bins centred at $z=4.2$, 4.6 and 5.0. The flux power spectrum uncertainties are lower by up to 40\%, and  extend to twice as small a scale, as previously resolved in Lyman-$\alpha$ forest studies (e.g. \citet{Viel13}), reaching $k_{\rm max} = 0.2\;\skm$. These improvements are the result of a larger statistical sample and more rigorous modeling of the observational systematic effects. This data set has been used to set the tightest constraints on the nature of dark matter \citep{Garzilli2021,Rogers2021,2023PhRvD.108b3502V,Irsic2024,GarciaGallego2025,Liu2026,Mosbech2026}, primordial features in the cosmological initial conditions \citep{Pavicevic2025,GarciaGallego2026}, and thermal history of \HI\ reionization in the IGM \citep{Boera2019,2026PhRvR...8c2009G,Irsic2026}, thus motivating its use in the current study. 

\subsection{Likelihood priors}

The measurements and the simulated models are compared within a Bayesian likelihood approach, building on the work of \cite{Molaro2022,Molaro2023}, namely using a Monte Carlo Markov Chains (MCMC) sampler with a Gaussian likelihood and the full flux power spectrum covariance of \citet{Irsic2024}. For the likelihood analysis, independent priors from \cite{planck18} on cosmological parameters ($\sigma_8,n_s$) were used, and an additional thermal prior in $u_0-T_0$ plane was included, following \cite{Molaro2022}. Recently, it has been shown that this choice of thermal prior, to a good approximation, reflects a prior on $\tau_e$ as measured from the CMB \citep{GarciaGallego2026}. Unless otherwise stated, only a wide flat prior was used for $\tau_{\rm eff}(z_i)$ parameters in the range of $[0.3,1.7]\times \tau_{\rm eff}^{\rm B19}(z_i)$ around the measured value of \citet{Boera2019}. This is an uninformative prior on a parameter that is independently measured to better than 5\% precision \citep{2013MNRAS.430.2067B,bosman18,Bosman2022}.

\section{Results}
\label{sec:results}

The results of the data analysis are presented in this section and summarized in Table~\ref{tab:results}. The likelihood analysis combined all three observed redshift bins, with 48 data points across them, as described in Sec.~\ref{sec:data}. The base $\Lambda$CDM homogeneous model was constructed using a suite of Sherwood-Relics simulations described in Sec.~\ref{sec:sims} and consists of four thermal parameters per redshift bin ($\tau_{\rm eff},T_0,\gamma,u_0$) and two cosmology parameters ($\sigma_8,n_s$) that are redshift-independent. Because the cosmology parameters have a strong prior the total degrees of freedom for a homogeneous model is $\text{d.o.f.}=48-12=36$. The results with the baseline homogeneous model show good agreement with previously published analyses \citep{Boera2019,Villasenor2022apj,Irsic2024} and a $\chi^2/\text{d.o.f.} = 42.1/36 = 1.17$.

\begin{table}
    \centering
    \begin{tabular}{c | c | c }
    \toprule
    Model & $\Mmax$ & $\chi^2 / \text{d.o.f.}$ \\
    \midrule\midrule \addlinespace[5pt]
    homogeneous & - & 42.1 / 36 \\ \addlinespace[5pt] \midrule\addlinespace[5pt]
    $\Delta T_b=1\times 10^4\;\mathrm{K}$ \& K19 model 1 & $-24.9^{+3.0}_{-2.7}$ & 39.5 / 35 \\ 
    $\Delta T_b=2\times 10^4\;\mathrm{K}$ \& K19 model 1 & $-24.8^{+2.6}_{-1.5}\left.^{+3.2}_{-4.5}\right.$ & 36.6 / 35 \\ \addlinespace[5pt]
    $\Delta T_b=2\times 10^4\;\mathrm{K}$ \& K19 model 3 & $>-24.4$ & 35.7 / 35 \\ 
    \addlinespace[5pt]
    \bottomrule
    \end{tabular}
    \caption{The best-fit $\chi^2$ and parameter values for different analyses choices using the \citet{Boera2019} data. The models span different temperature boosts of a \HeIII\ region ($\Delta T_b$) and different quasar luminosity function models from \citet{Kulkarni2019}.}
    \label{tab:results}
\end{table}

\subsection{Model with redshift-independent $\Mmax$}
\label{subsec:fixedMmax}

\begin{figure*}
    \includegraphics[width=\textwidth]{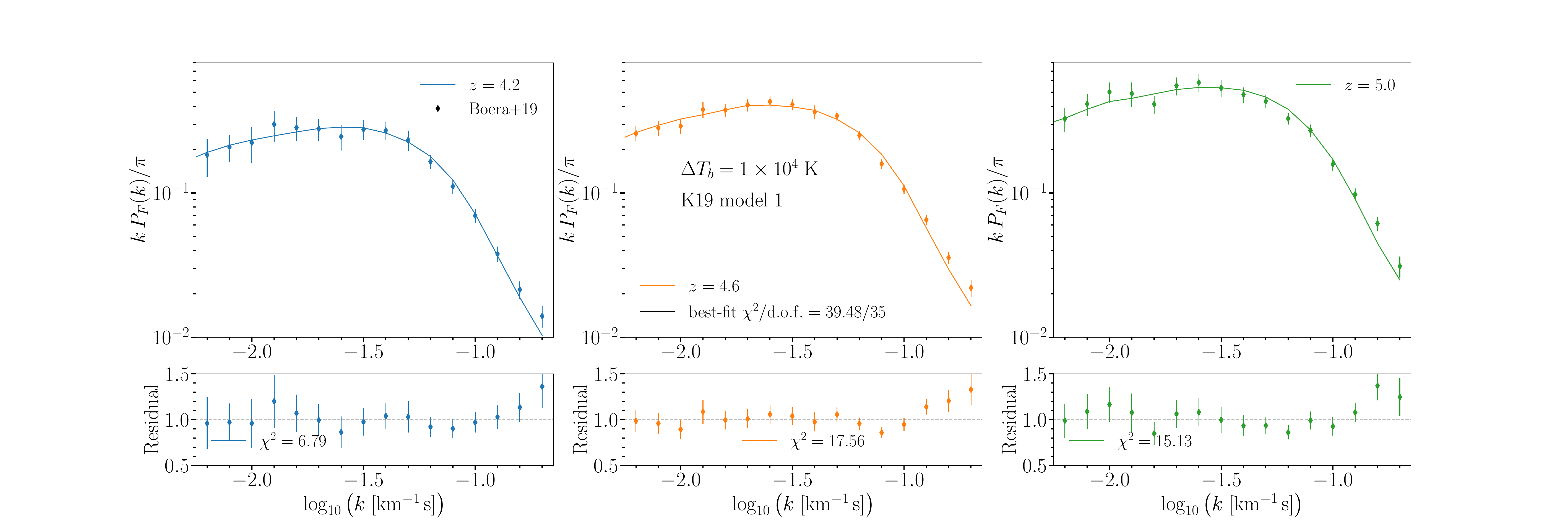}\\
	\includegraphics[width=\textwidth]{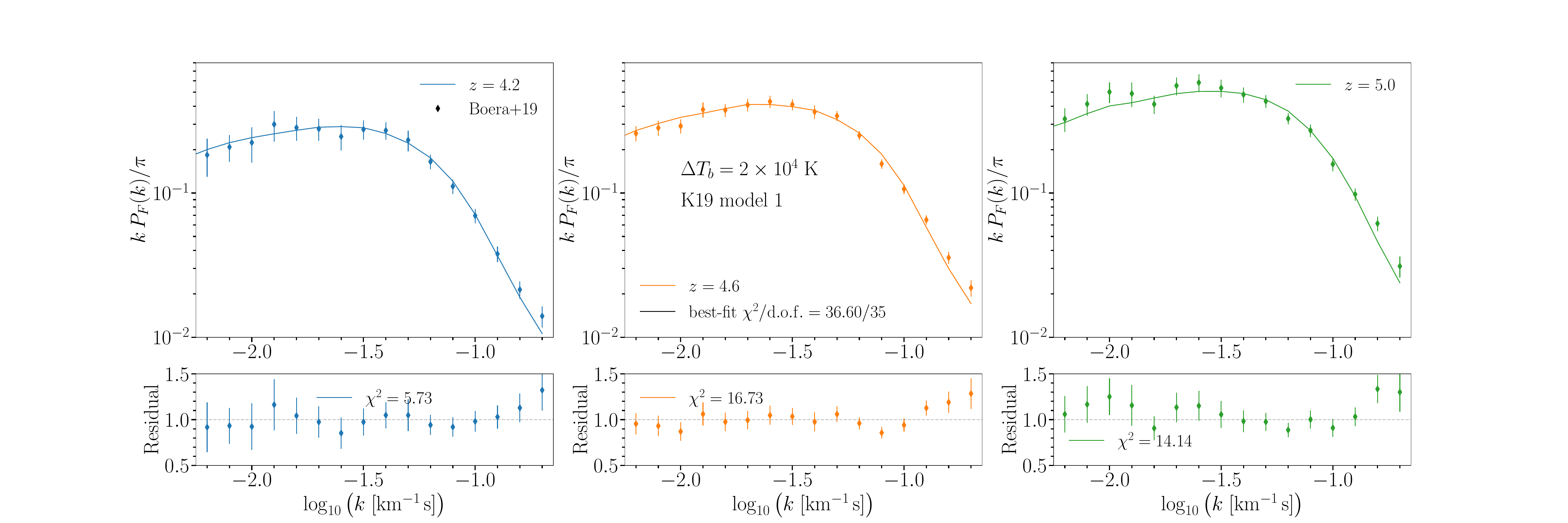}\\
    \includegraphics[width=\textwidth]{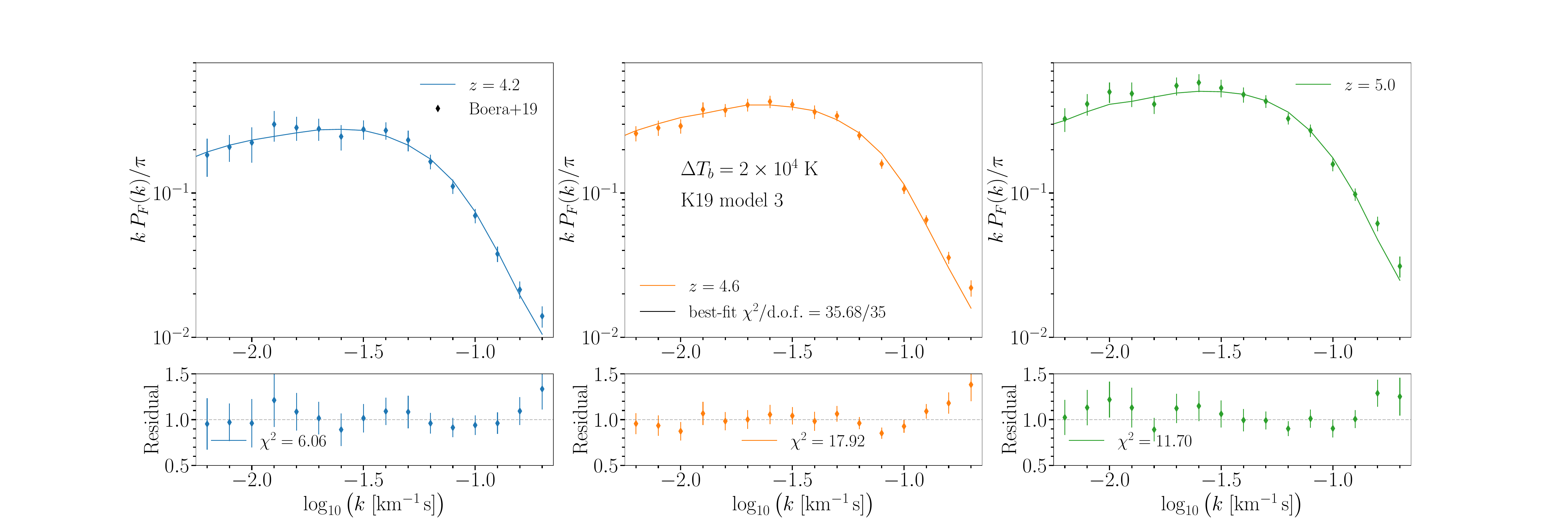}\\
    \caption{The best-fit $P_{\rm 1D}^{\rm QSO}$ model compared with the data of \citet{Boera2019}. Three different rows show the best fit for the three models from Table~\ref{tab:results}. In each row three panels correspond to three redshift bins, with the bottom panels showing the residuals of the data over the model. The data were compared to a simulation based model that varies three thermal parameters and mean transmission independently in each redshift bin ($\tau_{\rm eff}$ , $T_0$, $\gamma$, $u_0$), two cosmology parameters ($\sigma_8$,$n_s$) with Planck \citep{planck18} priors, and the maximum UV magnitude ($\Mmax$) of the \citet{Kulkarni2019} QSO luminosity function Model 1 ({\it top, middle}) or Model 3 ({\it bottom}) for the QSO assisted reionization models. A temperature boost of $\Delta T_b=1\times10^4$~K ({\it top}) or $\Delta T_b=2\times10^4$~K ({\it middle, bottom}) within the \HeIII\ regions has been assumed (see text for details). 
    }
    \label{fig:p1d_bestfit_Mmax}
\end{figure*}

\begin{figure}
	\includegraphics[width=\columnwidth]{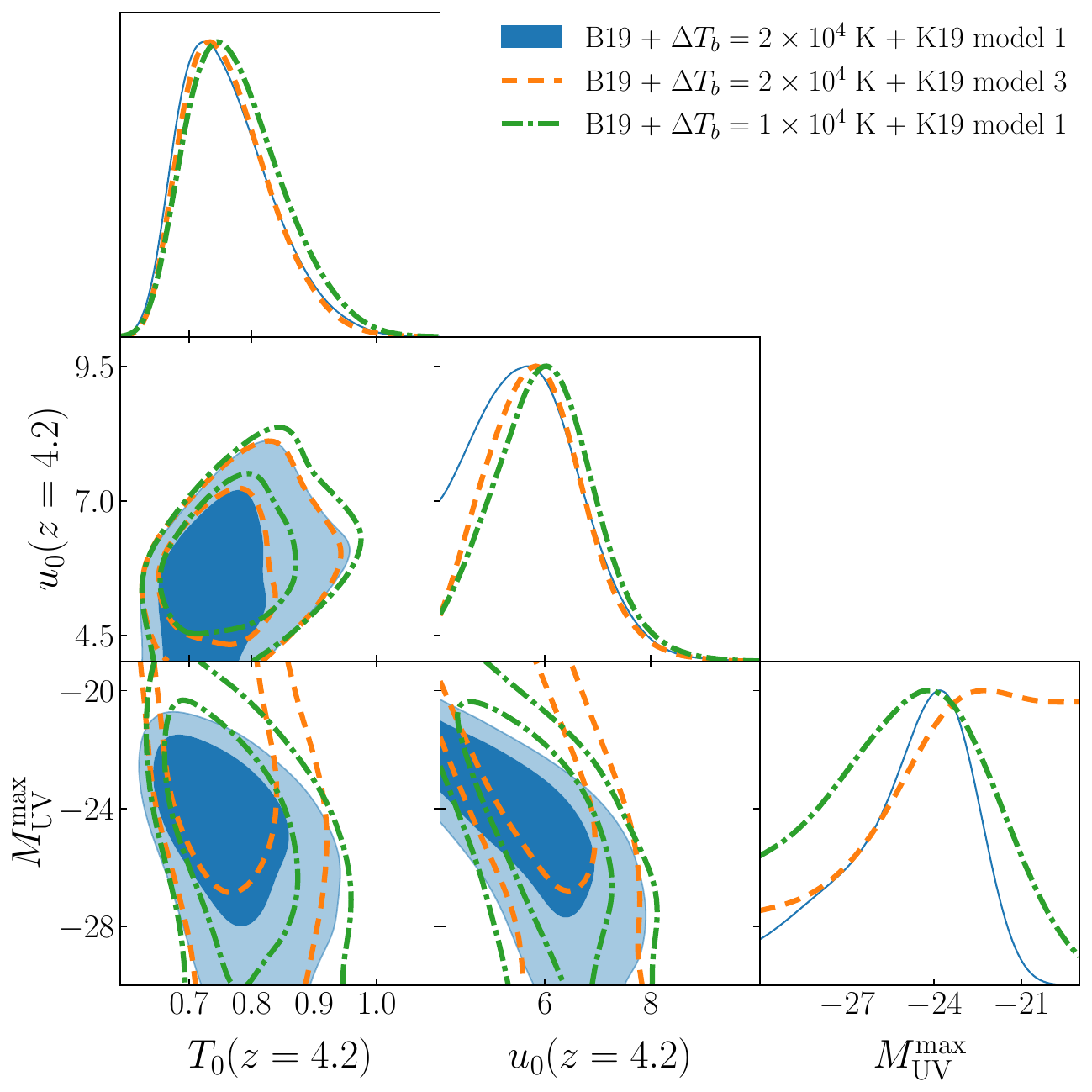}
    \caption{The 2D posterior probability distributions for the varied parameters in a likelihood analysis comparing QSO-assisted reionization models of Table~\ref{tab:results} with the P1D Lyman-$\alpha$ forest measurements from \citet{Boera2019}. The data were compared to a simulation based model that varies three thermal parameters and mean transmission independently in each redshift bin ($\tau_{\rm eff}$ , $T_0$, $\gamma$, $u_0$), two cosmology parameters ($\sigma_8$,$n_L$) with Planck \citep{planck18} priors, and maximum UV magnitude ($\Mmax$) of the QSO luminosity function for the QSO-assisted reionization models (see text for details). }
    \label{fig:panels_Mmax}
\end{figure}

The QSO assisted reionization models described in Sec.~\ref{sec:QSOmodels} are parametrized by the varying maximum magnitude $\Mmax$ that determined the integration limit on the faint end of the QSO luminosity functions of \citet{Kulkarni2019}. The QSO-assisted case using Model 1 and a temperature boost $\Delta T_b=1\times10^4$~K, corresponding to soft-spectra AGN, marginally improves the fits, shown in Fig.~\ref{fig:p1d_bestfit_Mmax}, with a favoured maximum redshift-independent magnitude upper limit $M_\mathrm{UV}^\mathrm{max}<-21.9$ ($1\sigma$), as shown in the lower right corner plot in Fig.~\ref{fig:panels_Mmax}, suggesting the dimmer AGN discovered by JWST do not contribute much to the ionization of \HeII. Our analysis, however, is not able to exclude an appreciable contribution from such a dim population with a high level of statistical confidence.

The results for an increased temperature boost of $\Delta T_b=2\times10^4$~K for Model 1 are shown in Fig.~\ref{fig:p1d_bestfit_Mmax}. The data again show a slight preference for a QSO-assisted model over the homogeneous baseline model, with $\Delta \chi^2 = -5.5$ for one additional degree of freedom. The best-fit value of $\Mmax = -24.8^{+2.6\,+3.2}_{-1.5\,-4.5}$ is consistent with no effect of QSO-assisted models ($\Mmax=-30$) at $\sim 2-3\sigma$. 
A contribution from faint QSOs with $M_\mathrm{UV}>-21.6$ is excluded at the $2\sigma$ level.
Most of the constraining power on the model comes from $z=4.2$ where the effect is largest (see Fig.~\ref{fig:p1d_qso_emulator}). 

The 2D posterior distribution for this redshift, shown in Fig.~\ref{fig:panels_Mmax}, reveals a strong degeneracy between two of the thermal parameters, $(T_0,u_0)$ that are mostly responsible for determining the small-scale 1D flux power spectrum suppression. (The 2D posterior distributions for the full set of fit parameters is shown in Appendix Fig.~\ref{fig:contour_Mmax}.) As the small-scale signal of the QSO-assisted models is also a suppression of the flux power at high wavenumbers, such a correlation among the recovered parameters is not unexpected. An analysis with QSO-assisted reionization leads to a colder IGM and with lower cumulative heat injection. While future observational constraints could potentially alleviate the issue of these degeneracies with thermal parameters by imposing independent priors, the risk is that the independent observational probes of the Doppler broadening, and thus $T_0$, might be dependent on the assumption of the level of QSO-assisted reionization. A similar argument may be made for the cumulative heat injection $u_0$, although informative priors may be derived more directly from the pressure smoothing scale \citep{Rorai2017,Irsic2026}, which shows low sensitivity to the nature of inhomogeneous reionization. However, this requires more detailed study using QSO-assisted reionization models.

Allowing for a large temperature boost $\Delta T_b=2\times10^4$~K for the QSO-assisted case with Model 3, corresponding to the number density of LRDs, does not much constrain the contribution of LRDs to \HeII\ ionization, with $\Mmax>-24.4$. This case, however, does give the greatest improvement in the fits in Fig.~\ref{fig:p1d_bestfit_Mmax}, with $\Delta\chi^2=-6.4$ for one additional degree of freedom.

\subsection{A model with varying $\Mmax$ as a function of redshift}
\label{subsec:Mmaxz}

The redshift-independent model of QSO-assisted reionization, however, is by construction model dependent as it inherits the redshift evolution of the underlying QSO luminosity function \citep{Kulkarni2019}. By adopting a parametrisation of the model where $\Mmax(z_i)$ is modelled as an independent free parameter in each of the observed redshift bins ($z_i \in [4.2,4.6,5.0]$) allows for the reconstruction of the redshift evolution of the effect. The most constraining redshift for the model is $z=4.2$ where the effect is strongest, favouring a contribution to \HeII\ ionization only from very bright QSOs, with $\Mmax<-24.5$ ($1\sigma$). The result at $z=4.6$ is consistent both with a contribution from bright QSOs with $\Mmax<-24.7\pm2.9$, or no effect from the QSOs at the $2\sigma$ level. It disfavours a contribution from dim AGN with $M_\mathrm{UV}>-21.8$. At $z=5.0$ the data prefer only a weak lower bound, with $\Mmax>-24.7$ ($1\sigma$).

The resulting posterior distributions for each of the three $\Mmax(z_i)$ parameters for QSO-assisted reionzation for Model 1 with $\Delta T_b=2\times10^4$~K are shown in Appendix Fig.~\ref{fig:contour_Mmax_zbin}.
The degeneracies found in the redshift-independent model are also present in the $\Mmax(z_i)$ analysis, with $\Mmax(z_i)$ at each redshift being degenerate with thermal parameters only at that redfshift, e.g. $u_0(z_i)$.

\subsection{Inference on ionizing emissivity}

Both of the models in the previous sections may be used to infer constraints on the amount of specific emissivity coming from the QSO population. The emissivity of QSOs at 912\AA\ may be used to estimate the contribution of QSOs to the reionization process. Under the assumption of a universal power-law SED of every quasar, $f_\nu \propto \nu^{-0.61}$ at $\lambda > 912$\AA, the specific emissivity is given by
\begin{equation}
    \epsilon_{\rm 912} = \left(\frac{912}{1450}\right)^{0.61}\,\int_{M_{\rm min}}^{\Mmax} \mathrm{d}M\, \phi(M,z)\, 10^{-0.4(M - 51.6)},
\end{equation}
where $M$ is the AB magnitude at 1450\AA\ and $\phi(M,z)$ is the QSO luminosity function. We give an estimate for Model 1 of \citet{Kulkarni2019}.

The posterior distribution from our analyses is mapped to $\epsilon_{\rm 912}$ at the level of the MCMC samples, and subsequently marginalized over the thermal parameters. For the redshift-independent $\Mmax$ model the emissivity due to QSOs increases by two decades over the redshift range of $z=7$ to $z=4$ \citep{Kulkarni2019}. The results are consistent with the analysis of a redshift-dependent $\Mmax(z_i)$. In both scenarios, the Lyman-$\alpha$ forest 1D flux power spectrum data limits the allowed contribution of QSOs to a specific emissivity below $\epsilon_{912} < 1-3 \times 10^{24}\;\mathrm{erg\,s^{-1}\,Hz^{-1}\,cMpc^{-3}} \;(2\sigma)$. Since the constraining power of the Lyman-$\alpha$ forest comes from the suppression of clustering on small-scales due to the QSO contribution to reionization, it is likely that the exact functional form of the QSO luminosity function to construct the models is not important; i.e., for a QSO luminosity function with higher amplitude compared with \citet{Kulkarni2019} Model 1, the data would provide more stringent constraints and limit the value of $\Mmax$ to only the bright-end for hard QSO spectra giving \HeIII\ regions a temperature boost of $\Delta T_b=2\times10^4$~K. Unlike the results of e.g., \citet{Kulkarni2019}, that are derived from the observed QSO luminosity function for a given range of observed $M_{\rm 1450}$, and require corrections for selection effects, the results from this study are derived from a different physical effect -- namely the role of the thermal heating following intergalactic \HeII\ photoionization by QSOs on the clustering of \HI\ in the IGM.

\begin{figure}
	\includegraphics[width=\columnwidth]{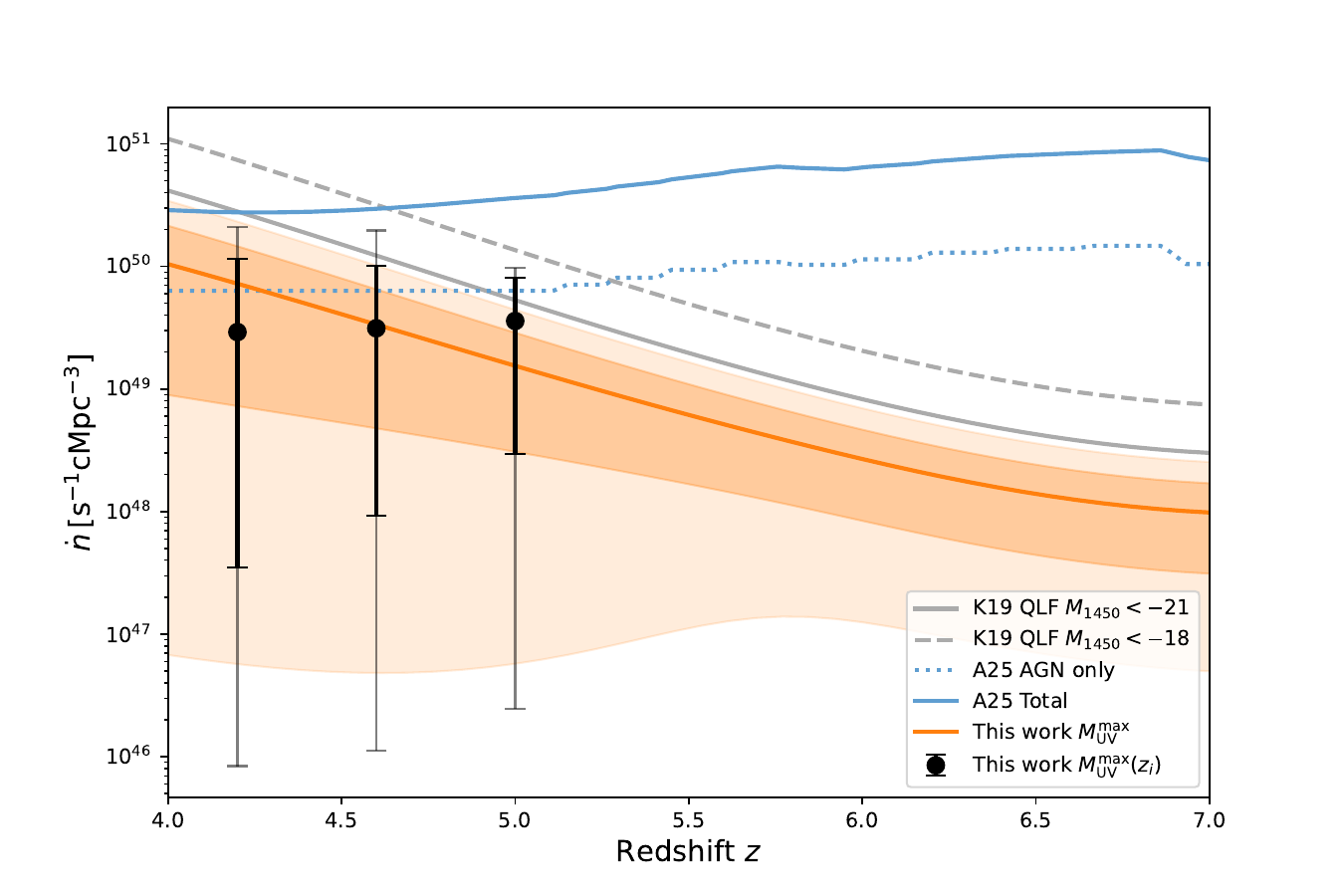}
    \caption{The total ionizing emissivity as derived from the analysis using the data from \citet{Boera2019}. The black data points show $1\sigma$ ($2\sigma$) results with the $M_{\rm max}^{\rm UV}(z_i)$ parametrisation that is independent for each observed redshift bin for QSO luminosity function Model 1 and temperature boost $\Delta T_b=2\times10^4$~K. The orange line and shaded regions show the corresponding median and $1\sigma$ ($2\sigma$) range when sampling only one global parameter $M_{\rm max}^{\rm UV}$ (including an extrapolation to $z>5$). The best-fit value for that parameter on the data is $M_{\rm max}^{\rm UV} = -24.8^{+2.6}_{-1.5}$. In comparison also shown are \citet{Kulkarni2019} models with QSO luminosity function contribution only from the bright end ($M_{1450}<-21$ solid gray) and including the faint end ($M_{1450}<-18$ dashed gray). The results are also compared to the \citet{Asthana2025b} QSO-assisted model (total in solid; AGN contribution only in dotted) that provide a good fit to the Lyman-$\alpha$ forest optical depth distribution at $z>5$.
    }
    \label{fig:k19_emissivity}
\end{figure}

 The specific emissivity results of our study can be mapped onto the total ionizing emissivity of \citet{Asthana2025} by adopting a constant escape fraction of QSOs $f_{\rm esc}=1$ and frequency dependence of the specific emissivity as $\epsilon_\nu \propto \nu^{\alpha_{\rm ion}}$, with $\alpha_{\rm ion}=-1.7$. The results are shown in Fig.~\ref{fig:k19_emissivity} for both $\Mmax$ and $\Mmax(z_i)$ models presented in this study. The results derived in this study constrain the amount of patchiness due to \HeIII\ temperature inhomogeniety resulting from the QSO population that assists the reionization, and are complementary to the analysis of \citet{Asthana2025b} which depends on the Lyman-$\alpha$ opacity fluctuations due to \HI\ reionization. The fact that the two approaches yield consistent results in the range of $z=4-5$ is encouraging. Fig.~\ref{fig:k19_emissivity} further highlights that the contribution from the bright end of the QSO luminosity function is in agreement with the results of this work, and that the contribution from faint QSOs with hard spectra has to be limited.

\subsection{Mock data analysis}
\label{subsec:mock_analysis}

In order to assess the feasibility of future survey data to constrain such QSO-assisted models the analysis based on Sec.~\ref{sec:results} was performed on mock data. 

We consider two sets of mock analyses to validate the inference framework. In the baseline mock, \HeII\ reionization is homogeneous, allowing us to evaluate how effectively a QSO-assisted scenario can be excluded. In the second setup, we model QSO-assisted \HeII\ reionization with a specified $\Mmax=-21$ to assess how accurately this parameter may be recovered.

For the ground truth in both setups, we adopt the {\tt Homog-late} simulation from \citet{Molaro2022}, which belongs to the Sherwood-Relics suite \citep{Puchwein2023} but was excluded from the grid used to construct our likelihood emulator. This simulation implements a homogeneous UV background within a $40\;\mathrm{cMpc}/h$ box containing $2\times 2048^3$ dark matter and baryon particles. Its thermal history was explicitly matched to the median thermal history of the corresponding inhomogeneous model \citep{Molaro2022,Puchwein2023}, which in turn was tuned to match observational constraints \citep[for details see][]{Molaro2022}.

In our default mock runs, we adopt the exact same parameter priors used in the analysis of the observational dataset. Additionally, we perform a run with modified $\tau_{\rm eff}$ priors: here, the Gaussian prior is centered directly on the ground truth values of the simulation, while the standard deviations are fixed to the empirical uncertainties from independent $\tau_{\rm eff}$ measurements \citep{2013MNRAS.430.2067B, Bosman2022}. This illustrates the power of independent informative priors on this parameter in future observational studies.

The mock covariance matrix structure was obtained through bootstrapping the power spectrum of the lines of sight through the ground truth simulations. The final covariance was finally rescaled under the assumption of being dominated by statistical uncertainties and the variance scaling with the number $N$ of Lyman-$\alpha$ forest sightlines as $\propto 1/N$. More realistic mocks \citep[e.g.][]{2017MNRAS.466.4332I,Ma2026b} support this assumption in the current regime of observations, where $N\sim 10-20$. For each mock setup, two versions of the mocks were constructed with targeted relative flux power spectrum uncertainties of 10\% and 5\%, respectively. These numbers bracket the near future observational programmes already collecting data (e.g. GHOSTLY \citep{Artola2024}, EQUALS \citep{Berg2025}) at 10\% relative uncertainty, and planned future programmes \citep[e.g. WST, ][]{Mainieri2024} at below 5\%. As a point of reference, the current observational data set of \citet{Boera2019} that is used in this study has on average 15-20\% uncertainties on the 1D flux power spectrum in the redshift range of $z=4.2-5.0$. For the sake of simplicity the mock data set was constructed with the same redshift- and k- binning as the data of \citet{Boera2019}. This is likely a good assumption for surveys in the near future.

In the baseline mock setup, the ground truth included only a homogeneous UV background and therefore the targeted $\Mmax=-30$, corresponding to no QSO assistance in the \HeII\ reionization. Repeating the analysis and assessing the mock data with a QSO-assisted model yields improved constraining power on the QSO reionization models with $\Mmax < -26.0\,(2\sigma)$ and $\Mmax < -27.3\,(2\sigma)$ for 10\% and 5\% mocks, respectively. In terms of the significance of the deviation from the targeted value of $\Mmax=-30$, this is a factor of 1.5 improvement in $\Delta \Mmax = \Mmax(2\sigma)-\Mmax(\mathrm{target})$, for a factor of $\sqrt{2}$ improvement in the observational uncertainties. Such an analysis of course only serves to illustrate the significance of {\it excluding} the QSO-assisted model if it is not present in the data.

\begin{figure}
	\includegraphics[width=\columnwidth]{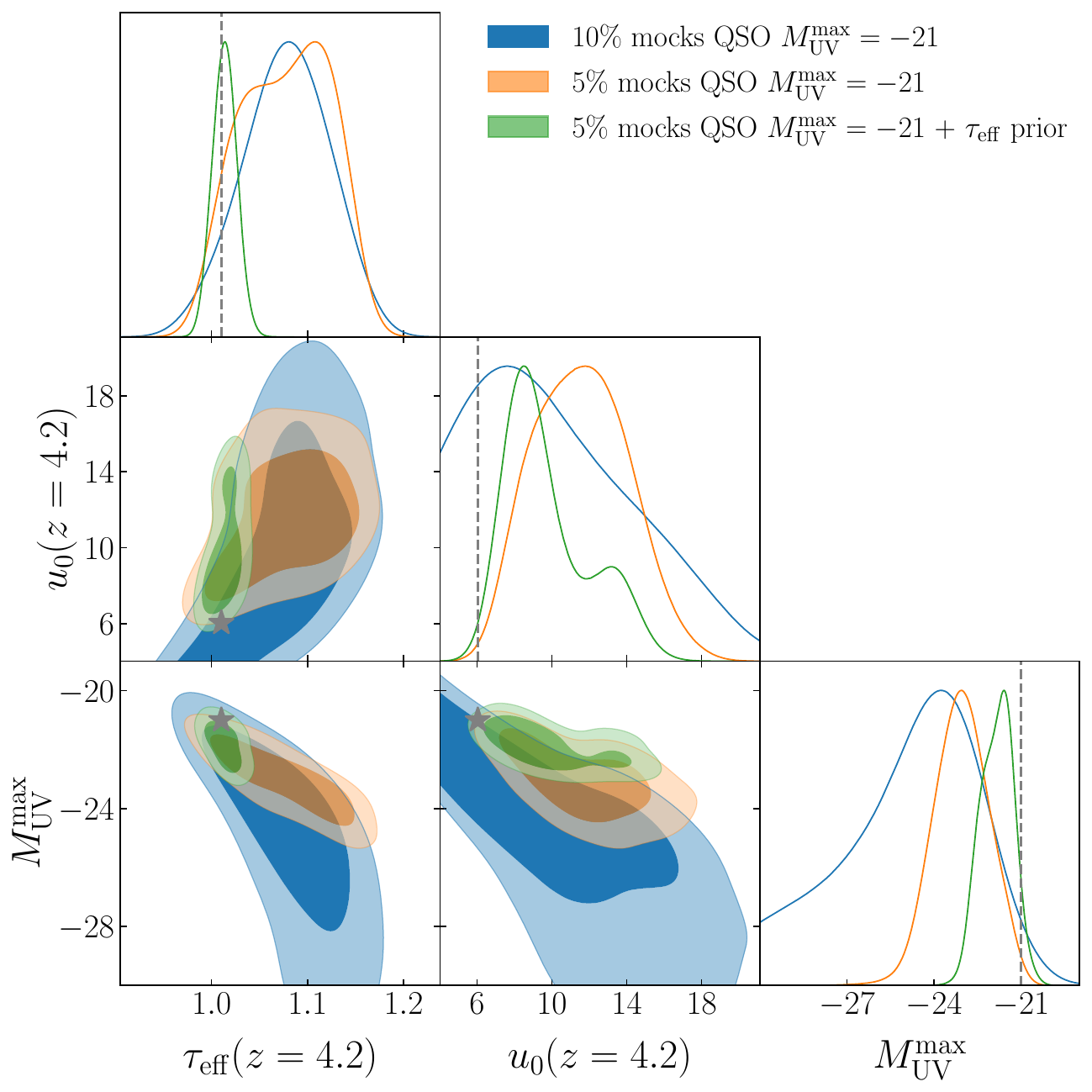}
    \caption{The 2D posterior probability distributions for the varied parameters in a likelihood analysis comparing different analyses of QSO-assisted reionization models for QSO-assisted mocks with a target $\Mmax = -21$. The mocks differ only in the relative uncertainty in the flux power spectrum, chosen to be 10\% and 5\%, in blue and orange, respectively. In green is the analysis on the mocks with 5\% uncertainty with added informative priors on $\tau_{\rm eff}$ from independent observations. The gray star/dashed line shows the true parameter values of the mocks.
    }
    \label{fig:panels_mocks_avery}
\end{figure}

The second mock setup was constructed for a QSO-assisted model with $M_\mathrm{UV}^\mathrm{max}=-21$ from Sec.~\ref{sec:QSOmodels} painted on top of the homogeneous mock flux power spectrum. The covariance was adjusted to reflect this change and to retain 10\% and 5\% relative uncertainty on the mock data. This setup was used to assess the significance with which a genuine presence of QSO assistance with a restricted $\Mmax$ in the reionization of \HeII\ is {\it detected} in future observational data. The results are shown in Fig.~\ref{fig:panels_mocks_avery}. The correlations between $\Mmax$ and thermal parameters (e.g. $u_0$) are found in the mock spectra analysis with similar levels of degeneracy as in the real data. As the uncertainty level in the flux power spectrum of the mock data decreases, a new correlation appears with the $\tau_{\rm eff}$ parameter to the point where it begins to dominate the correlation with $\Mmax$ in the 5\% mocks. This degeneracy axis is not present in the real data with higher $\sim 10-15\%$ flux uncertainties. 

The targeted value of $\Mmax=-21$ in the mocks was recovered within $1-2\sigma$ significance in both 10\% and 5\% mocks data, with best-fit values of $-24.7^{+2.8}_{-1.7}(1\sigma)^{+3.7}_{4.7}(2\sigma)$ and $-23.05^{+0.96}_{-0.96}(1\sigma)^{+1.8}_{-1.8}(2\sigma)$ respectively. The posterior distribution extends along the $\tau_{\rm eff} - \Mmax$ and $u_0-\Mmax$ degeneracy directions which, due to the size of the prior volumes, complicates the inference in the projected 1D $\Mmax$ confidence intervals. 

These results may be improved in future surveys through the use of informative priors. Unlike priors on thermal parameters which are often derived indirectly from observables that will likely be dependent on the choice of QSO-assisted reionization modelling, the informative priors on the mean transmission ($\tau_{\rm eff}$) are derived directly from observations and so agnostic to the modelling choice. With $\tau_{\rm eff}$ priors from \citet{2013MNRAS.430.2067B} (z=4.2,4.6) and \citet{Bosman2022} (z=5.0), the resulting constraint on QSO-assisted reionization models is $M_{\rm UV}^{\rm max} = -21.84^{+0.62}_{-0.55}(1\sigma)\left.^{+1.10}_{-1.10}(2\sigma)\right.$. Other model parameters also become somewhat more constrained, as shown in Appendix Fig.~\ref{fig:contour_mocks_avery}.

\section{Discsussion and Conclusions}
\label{sec:conclusions}

This work assesses the viability of QSO assisted reionization models from the perspective of the Lyman-$\alpha$ forest clustering data. The publicly available measurements of \citet{Boera2019} of the 1D Lyman-$\alpha$ forest flux power spectrum were compared to the likelihood emulator built on the Sherwood-Relics simulation suite \citep{Puchwein2023}. The QSO reionization modelling was parametrized by the faint end integration limit of the QSO luminosity function $\Mmax$. We considered two QSO luminosity function models, Models 1 and 3, from \citet{Kulkarni2019}. The faint end extrapolation of Model 1 corresponds to the number density of high redshift faint AGN recently discovered with JWST \citep{2023ApJ...959...39H, 2024A&A...691A.145M}. The faint end extrapolation of Model 3 corresponds to the much smaller sub-population of Little Red Dots \citep{2024Natur.628...57F, 2024ApJ...964...39G, 2024ApJ...963..129M}.

For Model 1 and a \HeIII\ region temperature boost of $\Delta T_b=1\times10^4$~K, corresponding to a soft QSO spectrum, the data show a slight preference for the QSO-assisted model with the restriction $\Mmax=-24.9^{+3.0}_{-2.7}$. At the $1\sigma$ level, this excludes a contribution from dim QSOs with $M_\mathrm{UV}>-21.9$, but within the limits of our anaysis allows for dimmer QSOs consistent with the population of faint AGN discovered by JWST.

For an increased \HeIII\ region temperature boost of $\Delta T_b=2\times10^4$~K for Model 1, corresponding to a hard QSO spectrum, the data show an even greater preference for the QSO-assisted model, but with the tighter restriction $\Mmax=-24.8^{+2.6 +3.2}_{-1.5 - 4.5}$. At the $2\sigma$ level, this excludes a contribution from dim AGN with $M_\mathrm{UV}>-21.6$ if they have hard spectra. A further study that treated $\Mmax(z_i)$ independently in each of the three observed redshifts $z=4.2, 4.6$ and $5.0$ confirmed this result. This suggests the population of dim AGN with $-21<M_\mathrm{UV}<-18$ discovered by JWST do not have hard spectra, consistent with their nearly complete non-detection in x-ray emission \citep{2025MNRAS.538.1921M}. Alternatively, the ionizing photon escape fraction of hard-spectra QSOs may be magnitude-dependent, with the faint end $M_{\rm 1450} > -21$ escape fraction values much lower ($f_{\rm esc} \ll 1$) than the bright end, thus reducing their capacity to emit ionizing photons and contribute to either \HI\ or \HeII\ reionization.

For QSO luminosity function Model 3, no strong restriction on $\Mmax$ for the $\Delta T_b=2\times10^4$~K case is found in the analysis presented in this work ($\Mmax > -24.4$), allowing the population of Little Red Dots
to contribute to \HeII\ reionization, and by implication, to \HI\ reionization as well, even if they have hard spectra.

The results highlight the degeneracy between the thermal parameters of the IGM, e.g. the cumulative heat injected during reionization, and the parameter of the QSO-assisted model. While these degeneracies could be broken using informative priors on the thermal history of the IGM, it will be up to future work to carefully consider whether such priors are indeed independent. It will also be important to consider the effects of QSO-driven temperature fluctuations together with the patchiness induced by the galaxy-driven \HI\ reionization, both of which may have an impact on the small-scale 1D flux power spectrum. Although current models suggest that the inhomogeneous effect of \HI\ reionization on small-scales is small compared with the QSO-assisted models investigated in this work \citep{Onorbe2019,Keating2019,Molaro2022}, it necessitates further study \citep{Wu2021,Cain2024,Etezad2026}. 

Mock analyses were performed to assess the viability of future observing capabilities to constrain these QSO-assisted models. The mocks consisted of ground truth parameters around homogeneous reionization as well as around QSO-assisted models, showing that even near future surveys \citep[GHOSTLY][]{Artola2024} \citep[EQUALS][]{Berg2025} will be in a position to provide improved constraining power on the QSO contribution to \HI\ reionization. The mock analysis with 5\% relative uncertainty on the 1D flux power spectrum also reveals a new degeneracy with the mean transmission ($\exp(-\tau_{\rm eff})$), emerges as the dominant source of uncertainty in the inferred $\Mmax$ posterior distributions. However, measurements of $\tau_{\rm eff}$ do not rely on the modelling of either \HI\ or \HeII\ reionization, so they may thus be safely applied as informative external priors in future analyses. A mock analysis shows that $\Mmax$ may be recovered with tight error constraints when allowing for external priors on $\tau_\mathrm{eff}$.

A slight preference for QSO-assisted models derived in this work stems from the \HeIII\ thermal inhomogeneities and their impact on small-scale 1D clustering of the Lyman-$\alpha$ forest. This is in line with recent work suggesting that the Lyman-$\alpha$ optical depth distribution of \citet{Bosman2022} may be better accounted for allowing for a modest contribution of low luminosity QSOs to the \HI\ cosmic reionization process \citep{2024MNRAS.535.1035M, Asthana2025b}. The two complementary approaches bound the contribution of QSOs to the ionizing emissivity, with the Lyman-$\alpha$ optical depth cumulative distribution function requiring some contribution of faint QSOs to \HI\ reionization \citep{2024MNRAS.535.1035M,
Asthana2025b}, and this work suggesting that too high a contribution from hard spectra QSOs is not able to explain the clustering data of the Lyman-$\alpha$ forest because of their boost to the IGM temperature in \HeIII\ regions. The restriction on the maximum UV magnitude for a temperature boost $\Delta T_b=2\times10^4$~K provides a very clear and robust way forward for future observations to constrain the hard-spectra QSO contribution to \HI\ reionization. Improved measurements of the \Lya\ forest power spectrum may restrict the magnitude range of soft-spectra QSOs contributing to \HeII\ ionization as well, and so the total contribution of QSOs to the hydrogen-ionizing photon budget if the escape fractions of \HeII-ionizing and \HI-ionizing photons are the same.

\section*{Acknowledgements}
The authors thank Shikhar Asthana for useful discussions. VI acknowledges support from the Higgs Centre for Theoretical Physics at the University of Edinburgh where parts of this work were completed. AM and VI also thank INAF-OATS and IFPU in Trieste for their support and hospitality. For the purpose of open access, the authors have applied a Creative Commons Attribution (CC BY) licence to any Author Accepted Manuscript version arising from this submission.

The simulations used in this work were performed using the Joliot Curie supercomputer at the Tré Grand Centre de Calcul (TGCC) and the Cambridge Service for Data Driven Discovery (CSD3), part of which is operated by the University of Cambridge Research Computing on behalf of the STFC DiRAC HPC Facility (www.dirac.ac.uk). We acknowledge the Partnership for Advanced Computing in Europe (PRACE) for awarding us time on Joliot Curie in the 16th call. The DiRAC component of CSD3 was funded by BEIS capital funding via STFC capital grants ST/P002307/1 and ST/R002452/1 and STFC operations grant ST/R00689X/1. This work also used the DiRAC@Durham facility managed by the Institute for Computational Cosmology on behalf of the STFC DiRAC HPC Facility. The equipment was funded by BEIS capital funding via STFC capital grants ST/P002293/1 and ST/R002371/1, Durham University and STFC operations grant ST/R000832/1. DiRAC is part of the National e-Infrastructure.

\section*{Data Availability}
The data and analysis code used in this work are available from the authors on request.  Further guidance for accessing  the publicly available Sherwood-Relics simulation data can be found on the project website:  \url{https://www.nottingham.ac.uk/astronomy/sherwood-relics/}



\bibliographystyle{mnras}
\bibliography{references} 




\appendix

\section{Full posterior distributions}

In this Appendix we present the full posterior distributions for all the parameters varied in the Bayesian likelihood analysis of Sec.~\ref{subsec:fixedMmax} and ~\ref{subsec:Mmaxz}.

The analysis used observational measurements of the Lyman-$\alpha$ flux P1D of \citet{Boera2019} and theoretical models based on Sherwood-Relics simulations. The analysis fit for all three observed redshift bins ($z=4.2, 4.6, 5.0$) simulateneously, with four astrophysical model parameters per redshift bin $(\tau_{\rm eff},T_0,\gamma,u_0)$. The QSO-assisted model parameters were either $\Mmax$ with fixed redshift dependent effect inherited from the redshift dependence of the \citet{Kulkarni2019} Model 1 Quasar Luminosity Function (QLF), or three redshift independent parameters $\Mmax(z_i)$, for each redshift bin respectively. The specific QSO-assisted model constrained in this case is $\Delta T_b=2\times 10^4$~K from Table~\ref{tab:results}.

\begin{figure*}
	\includegraphics[width=\textwidth]{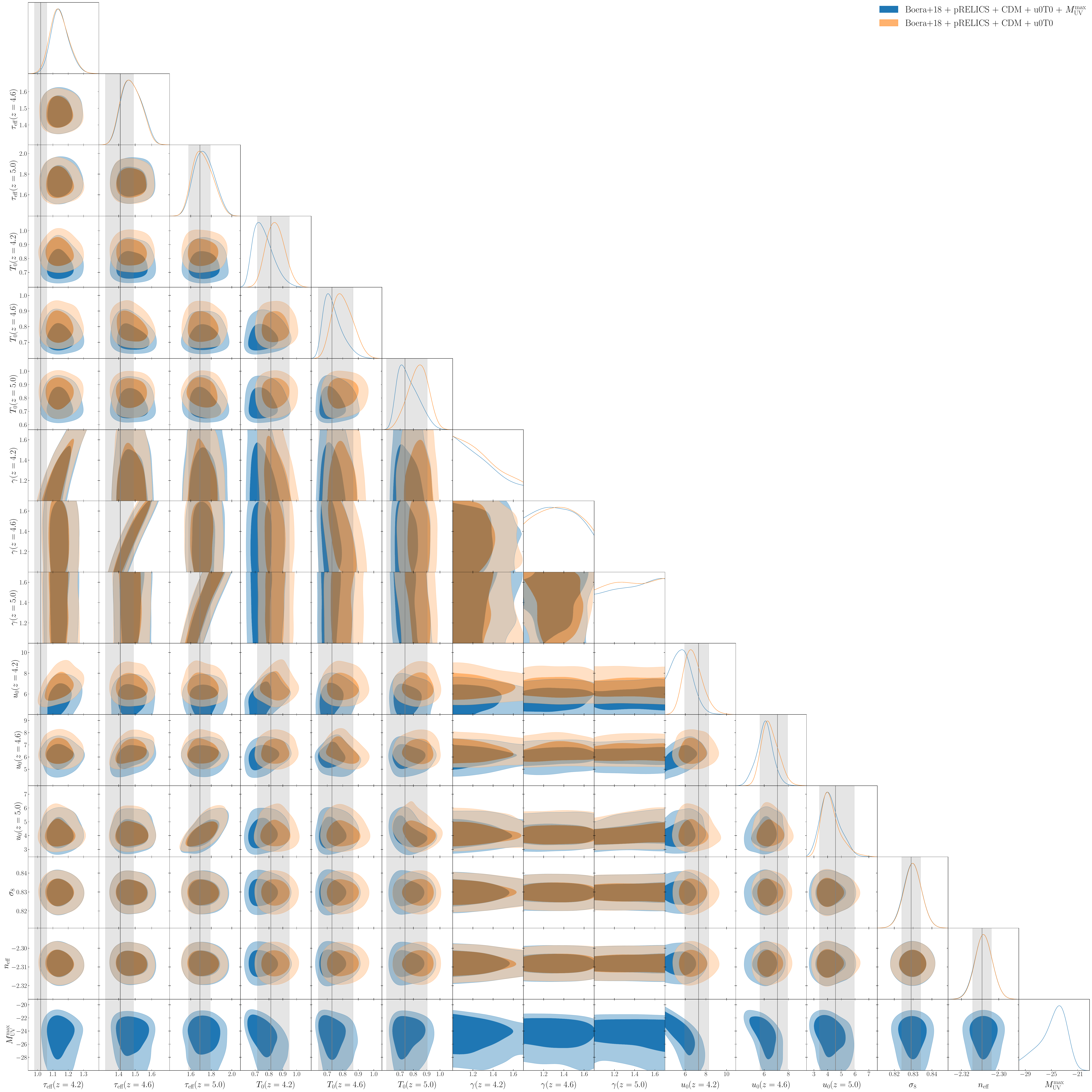}
    \caption{The 2D posterior probability distributions for the varied parameters in a likelihood analysis comparing QSO-assisted reionization models for a \HeIII\ region temperature boost $\Delta T_b=2\times10^4$~K for \citet{Kulkarni2019} Model 1, with the P1D Lyman-$\alpha$ forest measurements of \citet{Boera2019}. The data were compared to a simulation based model that varies three thermal parameters and mean transmission independently in each redshift bin ($\tau_{\rm eff}$ , $T_0$, $\gamma$, $u_0$), two cosmology parameters ($\sigma_8$,$n_L$) with Planck \citep{planck18} priors, and maximum UV magnitude ($\Mmax$) of the QSO luminosity function for the QSO-assisted reionization models (see text for details). }
    \label{fig:contour_Mmax}
\end{figure*}

Fig.~\ref{fig:contour_Mmax} shows the 2D posterior distributions for all the varied parameters. In blue are shown the results of the QSO-assisted analysis, and in orange are shown the results of a consistency test on the thermal history, where QSO-assisted model parameters were kept fixed to the values that result in a homogeneous UV background. The results show that while minor shifts in the thermal parameters $T_0$ and $u_0$ are seen between the two analyses, the  resulting thermal history is consistent after the inclusion of QSO-assisted model parameters.

At the level of the relative P1D uncertainty in the data \citep{Boera2019}, the dominant degeneracy direction is between $\Mmax$ and $(T_0,u_0)$ combination, in such a way that higher values of $\Mmax$ push the thermal history into the regime of lower temperature and lower cumulative heat injection. These findings are consistent with the results of \citep{Molaro2023,Irsic2024} where inhomogeneous \HI\ reionzation models show a similar effect.

\begin{figure*}
	\includegraphics[width=\textwidth]{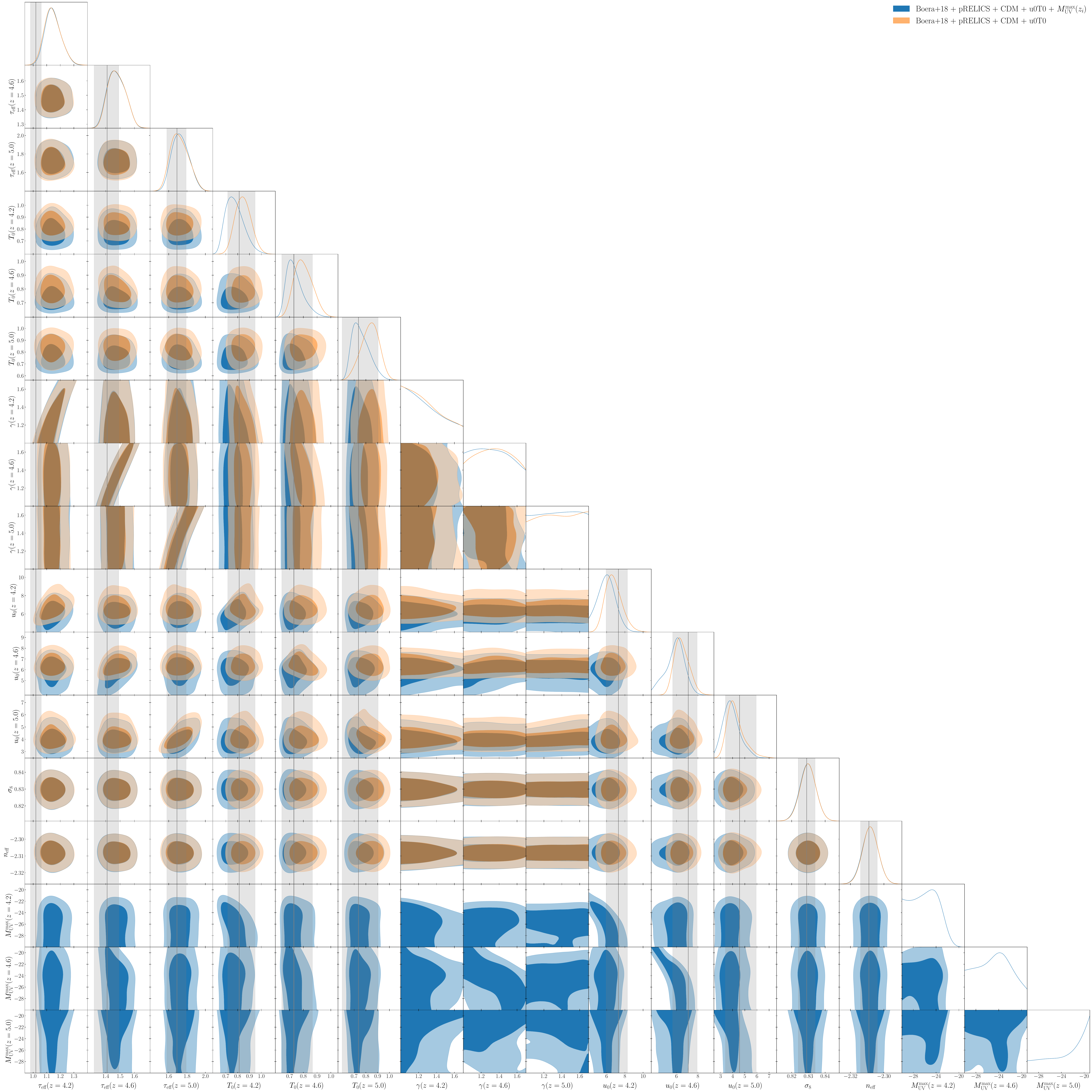}
    \caption{The 2D posterior probability distributions for the varied parameters in a likelihood analysis comparing QSO-assisted reionization models for a \HeIII\ region temperature boost $\Delta T_b=2\times10^4$~K for \citet{Kulkarni2019} Model 1, with the P1D Lyman-$\alpha$ forest measurements of \citet{Boera2019}. The data were compared to a simulation based model that varies three thermal parameters and mean transmission independently in each redshift bin ($\tau_{\rm eff}$ , $T_0$, $\gamma$, $u_0$), two cosmology parameters ($\sigma_8$,$n_L$) with Planck \citep{planck18} priors, and three maximum UV magnitudes ($\Mmax(z_i)$), one for each redshift bin of observations, of the QSO luminosity function for the QSO-assisted reionization models (see text for details).}
    \label{fig:contour_Mmax_zbin}
\end{figure*}

Fig.~\ref{fig:contour_Mmax_zbin} presents the results of the similar analysis, but with redshift independent $\Mmax(z_i)$ parametrisation. The resulting constraining power on $\Mmax(z_i)$ is weaker in this more conservative analysis where no redshift evolution of the effect is imposed. Only the middle of the redshift bins is effectively constrained by the data alone, while the lowest and highest redshift bins depend on the prior volume.

The dominant degeneracy axes and parameter combinations are preserved, although the degeneracy is now much stronger between the corresponding $(\Mmax,T_0,u_0)$ parameters at lower redshift ($z=4.2$) and almost entirely disappears at higher redshift. This is a reflection of the fact that the QSO-assisted model ($\Delta T_b=2\times 10^4$~K, Model 1 \citet{Kulkarni2019} QLF; see Table~\ref{tab:results}) has strongest impact on the Lyman-$\alpha$ P1D at $z=4.2$.

\begin{figure*}
	\includegraphics[width=\textwidth]{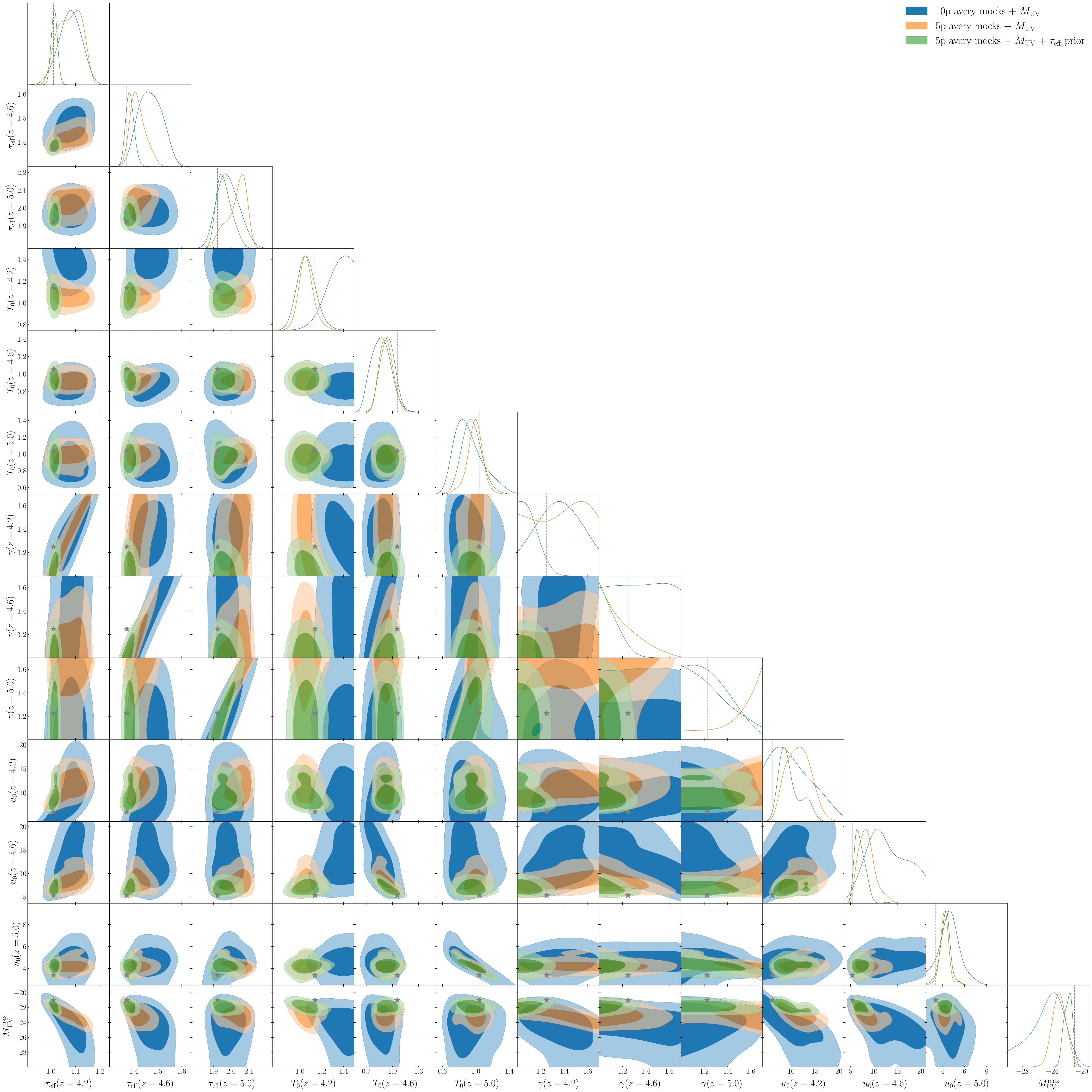}
    \caption{The 2D posteriors for the varied parameter in a likelihood analysis comparing QSO-assisted reionization models to the P1D Lyman-$\alpha$ forest mocks that include $M_{\rm UV}^{\rm target}=-21$. Measurements from QSO assisted mocks with 10\% and 5\% relative uncertainty on the flux power give $M_{\rm UV}=-24.7_{-1.7}^{+2.8}(1\sigma)\left._{-4.7}^{+3.7}\right.(2\sigma)$ and $M_{\rm UV} = -23.05_{-0.96}^{+0.96}(1\sigma)\left._{-1.8}^{+1.8}(2\sigma)\right.$, respectively. When an additional informative $\tau_{\rm eff}$ prior is used in combination with the 5\% mocks, the target value is recovered with $\Mmax=-21.84^{+0.62}_{-0.55}(1\sigma)\left._{-1.1}^{+1.1}(2\sigma)\right.$.}
    
    \label{fig:contour_mocks_avery}
\end{figure*}

The last figure, Fig.~\ref{fig:contour_mocks_avery}, show the full 2D posterior distributions for three mock analyses from Sec.~\ref{subsec:mock_analysis}. In all three cases the mocks consisted of a QSO-assisted model with $\Mmax=-21$ using Model 1 of \citet{Kulkarni2019} QLF with $\Delta T_b=2\times10^4$~K. The analyses shown are for 10\% relative uncertainty on mock P1D (blue), 5\% uncertainty (orange), and 5\% uncertainty with additional informative prior on $\tau_{\rm eff}$ (green).

As the uncertainty on P1D decreases, new degeneracies emerge between the thermal parameters and $\Mmax$. This is especially true at $z=4.2$ where the P1D shows strongest sensitivity to the QSO-assisted model parameter. Because the true value of $\Mmax$ in the mocks is close to the prior boundary, together with strong degeneracies in the $\Mmax-u_0$ and $\Mmax-\tau_{\rm eff}$ directions, prior volume effects arise where the 1D posterior distribution of $\Mmax$ is not centered on the true underlying value in the mocks.

The strong degeneracies with $u_0$ are difficult to break, as the measurements on the pressure smoothing scale, of which $u_0$ is a proxy, are typically performed precisely using P1D data \citep[although see][for alternatives]{rorai13,Irsic2026}. Independent measurements of $\tau_{\rm eff}$ do exist, however, and are agnostic to the modelling of the thermal and reionization history as they are derived directly from averaging the observed flux. The 2D posteriors in Fig.~\ref{fig:contour_mocks_avery} (green), show the effect of using informative priors on $\tau_{\rm eff}$ using a Gaussian distribution centered on the true value of $\tau_{\rm eff}$ in the mocks, with uncertainty coming from the current generation of independent observations \citep{Becker13,Bosman2022}.

\bsp	
\label{lastpage}
\end{document}